\documentclass[conference,a4paper]{IEEEtran}
\usepackage{cite,multirow}
\usepackage{morefloats}
\usepackage[acronyms,nonumberlist,nopostdot,nomain,nogroupskip]{glossaries}
\usepackage{soul}
\usepackage{amsmath,bbm}
\usepackage{amsfonts,amssymb,graphicx,subfigure,tikz}
\usepackage{etoolbox}
\usepackage{color}
\usepackage{diagbox}
\usepackage{epstopdf}
\usepackage{enumerate}
\usepackage{siunitx}
\usepackage{mathtools}
\usepackage{tikz}
\usepackage[utf8]{inputenc}
\usepackage{pgfplots} 
\usepackage{pgfgantt}
\usepackage{pdflscape}
\usepackage{comment}
\usepackage{bm}
\usetikzlibrary{arrows, positioning, patterns, patterns.meta}
\pgfplotsset{compat=newest}
\pgfplotsset{plot coordinates/math parser=false}
\usepgfplotslibrary{patchplots,groupplots}
\newlength\fheight
\newlength\fwidth
\newacronym{ca}{CA}{Carrier Aggregation}
\newacronym{3gpp}{3GPP}{3rd Generation Partnership Project}
\newacronym{5g}{5G}{5th generation}
\newacronym{5gc}{5GC}{5G Core}
\newacronym{adc}{ADC}{Analog to Digital Converter}
\newacronym{afbw}{AFBW}{Average Fading Bandwidth}
\newacronym{aimd}{AIMD}{Additive Increase Multiplicative Decrease}
\newacronym{am}{AM}{Acknowledged Mode}
\newacronym{amc}{AMC}{Adaptive Modulation and Coding}
\newacronym{aoa}{AoA}{Angle of Arrival}
\newacronym{aod}{AoD}{Angle of Departure}
\newacronym{ap}{AP}{Access Point}
\newacronym{aqm}{AQM}{Active Queue Management}
\newacronym{awgn}{AGWN}{Additive White Gaussian Noise}
\newacronym{balia}{BALIA}{Balanced Link Adaptation}
\newacronym{bdp}{BDP}{Bandwidth-Delay Product}
\newacronym{ber}{BER}{Bit Error Rate}
\newacronym{bf}{BF}{Beamforming}
\newacronym{bwp}{BWP}{Bandwidth Part}
\newacronym{cad}{CAD}{Computer-Aided Design}
\newacronym{cc}{CC}{Congestion Control}
\newacronym{cdf}{CDF}{Cumulative Distribution Function}
\newacronym{cir}{CIR}{Channel Impulse Response}
\newacronym{cn}{CN}{Core Network}
\newacronym{cp}{CP}{Control Plane}
\newacronym{cqi}{CQI}{Channel Quality Information}
\newacronym{crs}{CRS}{Cell Reference Signal}
\newacronym{csirs}{CSI-RS}{Channel State Information - Reference Signal}
\newacronym{dc}{DC}{Dual Connectivity}
\newacronym{dce}{DCE}{Direct Code Execution}
\newacronym{dci}{DCI}{Downlink Control Information}
\newacronym{dl}{DL}{Downlink}
\newacronym{dmr}{DMR}{Deadline Miss Ratio}
\newacronym{dmrs}{DMRS}{DeModulation Reference Signal}
\newacronym{dray}{D-Ray}{Deterministic Ray}
\newacronym{e2e}{E2E}{End-to-End}
\newacronym{ecn}{ECN}{Explicit Congestion Notification}
\newacronym{edf}{EDF}{Earliest Deadline First}
\newacronym{enb}{eNB}{evolved Node Base}
\newacronym{endc}{EN-DC}{E-UTRAN-\gls{nr} \gls{dc}}
\newacronym{epc}{EPC}{Evolved Packet Core}
\newacronym{es}{ES}{Edge Server}
\newacronym{fdd}{FDD}{Frequency Division Duplexing}
\newacronym{fdma}{FDMA}{Frequency Division Multiple Access}
\newacronym{fray}{F-Ray}{Flashing Ray}
\newacronym{fs}{FS}{Fast Switching}
\newacronym{ftp}{FTP}{File Transfer Protocol}
\newacronym{gmm}{GMM}{Gaussian Mixture Model}
\newacronym{gnb}{gNB}{Next Generation Node Base}
\newacronym{harq}{HARQ}{Hybrid Automatic Repeat reQuest}
\newacronym{hetnet}{HetNet}{Heterogeneous Network}
\newacronym{hh}{HH}{Hard Handover}
\newacronym{hol}{HOL}{Head-of-Line}
\newacronym{hqf}{HQF}{Highest-quality-first}
\newacronym{ia}{IA}{Initial Access}
\newacronym{iab}{IAB}{Integrated Access and Backhaul}
\newacronym{imt}{IMT}{International Mobile Telecommunication}
\newacronym{inf}{InF}{Indoor Factory}
\newacronym{inf-sh}{InF-SH}{\gls{inf}-Sparse High}
\newacronym{inr}{INR}{Interference to Noise Ratio}
\newacronym{iot}{IoT}{Internet of Things}
\newacronym{ked}{KED}{Knife-Edge Diffraction}
\newacronym{kpi}{KPI}{Key Performance Indicator}
\newacronym{lcf}{LCF}{Level Crossing Frequency}
\newacronym{lcr}{LCR}{Level Crossing Rate}
\newacronym{los}{LoS}{Line-of-Sight}
\newacronym{lte}{LTE}{Long Term Evolution}
\newacronym{ltemtp}{LTE-M}{LTE-MTC [Machine Type Communication]}
\newacronym{m2m}{M2M}{Machine to Machine}
\newacronym{mac}{MAC}{Medium Access Control}
\newacronym{mc}{MC}{Multi-Connectivity}
\newacronym{mcs}{MCS}{Modulation and Coding Scheme}
\newacronym{mec}{MEC}{Mobile Edge Cloud}
\newacronym{mi}{MI}{Mutual Information}
\newacronym{mib}{MIB}{Master Information Block}
\newacronym{mimo}{MIMO}{Multiple Input Multiple Output}
\newacronym{m-mimo}{m-MIMO}{massive-MIMO}
\newacronym{mlr}{MLR}{Maximum-local-rate}
\newacronym{mmwave}{mmWave}{millimeter wave}
\newacronym{moi}{MoI}{Method of Images}
\newacronym{mpc}{MPC}{Multi Path Component}
\newacronym{mptcp}{MPTCP}{Multipath TCP}
\newacronym{mr}{MR}{Maximum Rate}
\newacronym{mrdc}{MR-DC}{Multi \gls{rat} \gls{dc}}
\newacronym{mss}{MSS}{Maximum Segment Size}
\newacronym{mtd}{MTD}{Machine-Type Device}
\newacronym{mtu}{MTU}{Maximum Transmission Unit}
\newacronym{nfv}{NFV}{Network Function Virtualization}
\newacronym{nist}{NIST}{National Institute of Standards and Technology}
\newacronym{nlos}{NLoS}{Non-Line-of-Sight}
\newacronym{nr}{NR}{New Radio}
\newacronym{nrmse}{NRMSE}{Normalized Root Mean Square Error}
\newacronym{nsa}{NSA}{Non Stand Alone}
\newacronym{o2i}{O2I}{Outdoor-to-Indoor}
\newacronym{ofdm}{OFDM}{Orthogonal Frequency Division Multiplexing}
\newacronym{pa}{PA}{Position-aware}
\newacronym{prr}{PRR}{Packet Reception Ratio}
\newacronym{pbch}{PBCH}{Physical Broadcast Channel}
\newacronym{pdcch}{PDCCH}{Physical Downlonk Control Channel}
\newacronym{pdcp}{PDCP}{Packet Data Convergence Protocol}
\newacronym{pdsch}{PDSCH}{Physical Downlink Shared Channel}
\newacronym{pdu}{PDU}{Packet Data Unit}
\newacronym{per}{PER}{Packet Error Rate}
\newacronym{pf}{PF}{Proportional Fair}
\newacronym{pgw}{PGW}{Packet Gateway}
\newacronym{phy}{PHY}{Physical}
\newacronym{pl}{PL}{Path Loss}
\newacronym{ppp}{PPP}{Poisson Point Process}
\newacronym{prb}{PRB}{Physical Resource Block}
\newacronym{pss}{PSS}{Primary Synchronization Signal}
\newacronym{pucch}{PUCCH}{Physical Uplink Control Channel}
\newacronym{pusch}{PUSCH}{Physical Uplink Shared Channel}
\newacronym{qam}{QAM}{Quadrature Amplitude Modulation}
\newacronym{qd}{QD}{Quasi Deterministic}
\newacronym{rach}{RACH}{Random Access Channel}
\newacronym{ran}{RAN}{Radio Access Network}
\newacronym[firstplural=Radio Access Technologies (RATs)]{rat}{RAT}{Radio Access Technology}
\newacronym{red}{RED}{Random Early Detection}
\newacronym{rf}{RF}{Radio Frequency}
\newacronym{rlc}{RLC}{Radio Link Control}
\newacronym{rlf}{RLF}{Radio Link Failure}
\newacronym{rr}{RR}{Round Robin}
\newacronym{rray}{R-Ray}{Random Ray}
\newacronym{rrc}{RRC}{Radio Resource Control}
\newacronym{rrm}{RRM}{Radio Resource Management}
\newacronym{rs}{RS}{Remote Server}
\newacronym{rsrp}{RSRP}{Reference Signal Received Power}
\newacronym{rsrq}{RSRQ}{Reference Signal Received Quality}
\newacronym{rss}{RSS}{Received Signal Strength}
\newacronym{rssi}{RSSI}{Received Signal Strength Indicator}
\newacronym{rt}{RT}{Ray Tracer}
\newacronym{rtt}{RTT}{Round Trip Time}
\newacronym{rw}{RW}{Receive Window}
\newacronym{rx}{RX}{Receiver}
\newacronym{sa}{SA}{standalone}
\newacronym{sack}{SACK}{Selective Acknowledgment}
\newacronym{sap}{SAP}{Service Access Point}
\newacronym{sch}{SCH}{Secondary Cell Handover}
\newacronym{scm}{SCM}{Spatial Channel Model}
\newacronym{scoot}{SCOOT}{Split Cycle Offset Optimization Technique}
\newacronym{sdma}{SDMA}{Spatial Division Multiple Access}
\newacronym{sf}{SF}{Shadow Fading}
\newacronym{si}{SI}{Study Item}
\newacronym{sib}{SIB}{Secondary Information Block}
\newacronym{sinr}{SINR}{Signal-to-Interference-plus-Noise Ratio}
\newacronym{sir}{SIR}{Signal-to-Interference Ratio}
\newacronym{sm}{SM}{Saturation Mode}
\newacronym{snr}{SNR}{Signal-to-Noise Ratio}
\newacronym{son}{SON}{Self-Organizing Network}
\newacronym{srs}{SRS}{Sounding Reference Signal}
\newacronym{ss}{SS}{Synchronization Signal}
\newacronym{ssb}{SSB}{Synchronization Signal Block}
\newacronym{sss}{SSS}{Secondary Synchronization Signal}
\newacronym{sta}{STA}{Station}
\newacronym{tb}{TB}{Transport Block}
\newacronym{tcp}{TCP}{Transmission Control Protocol}
\newacronym{udp}{UDP}{User Datagram Protocol}
\newacronym{tdd}{TDD}{Time Division Duplexing}
\newacronym{tdma}{TDMA}{Time Division Multiple Access}
\newacronym{tfl}{TfL}{Transport for London}
\newacronym{tgad}{TGad}{Task Group ad}
\newacronym{tgay}{TGay}{Task Group ay}
\newacronym{tm}{TM}{Transparent Mode}
\newacronym{trp}{TRP}{Transmitter Receiver Pair}
\newacronym{tti}{TTI}{Transmission Time Interval}
\newacronym{ttt}{TTT}{Time-to-Trigger}
\newacronym{tx}{TX}{Transmitter}
\newacronym{ue}{UE}{User Equipment}
\newacronym{ul}{UL}{Uplink}
\newacronym{um}{UM}{Unacknowledged Mode}
\newacronym{uma}{UMa}{Urban Macro}
\newacronym{uml}{UML}{Unified Modeling Language}
\newacronym{utc}{UTC}{Urban Traffic Control}
\newacronym{vm}{VM}{Virtual Machine}
\newacronym{wbf}{WBF}{Wired Bias Function}
\newacronym{wf}{WF}{Wired-first}
\newacronym{wifi}{Wi-Fi}{Wireless Fidelity}
\newacronym{wigig}{WiGig}{Wireless Gigabit}
\newacronym{wlan}{WLAN}{Wireless Local Area Network}
\newacronym{xpr}{XPR}{Cross Polarization Ratio}
\newacronym{fr2}{FR2}{Frequency Range 2}
\newacronym{nbiot}{NB-IoT}{Narrowband-IoT}
\newacronym{cps}{CPS}{Cyber-Physical production System}
\newacronym{iiot}{IIoT}{Industrial Internet of Things}
\newacronym{agv}{AGV}{Autonomous Ground Vehicle}
\newacronym{uav}{UAV}{Unmanned Autonomous Vehicle}
\newacronym{amr}{AMR}{Autonomous Mobile Robots}
\newacronym{wsn}{WSN}{Wireless Sensor Network}
\newacronym{embb}{eMBB}{enhanced Mobile Broadband}
\newacronym{urllc}{URLLC}{Ultra-Reliable Low-Latency Communications}
\newacronym{lpwan}{LPWAN}{Low-Power Wide Area Network}
\newacronym{lora}{LoRa}{Long Range}
\newacronym{qos}{QoS}{Quality of Service}
\newacronym{ec}{EC}{energy consumption}
\usepackage[ruled,vlined]{algorithm2e}
\graphicspath{{Figures/}}

\begin{document}             
\title{Minimizing Energy Consumption for \\ 5G NR Beam Management for RedCap Devices}
 

\author{\IEEEauthorblockN{ Manishika~Rawat$^*$, Matteo~Pagin$^{*}$, Marco Giordani$^*$, Louis-Adrien Dufrene$^\dagger$, Quentin Lampin$^\dagger$, Michele Zorzi$^*$}
\IEEEauthorblockA{
$^*$Department of Information Engineering, University of Padova, Italy \\ 
$^\dagger$Orange Labs, France}}

\maketitle
\begin{abstract}
In 5G \gls{nr}, beam management entails periodic and continuous transmission and reception of control signals in the form of synchronization signal blocks (SSBs), used to perform initial access and/or channel estimation. 	
However, this procedure demands continuous energy consumption, which is particularly challenging to handle for low-cost, low-complexity, and battery-constrained devices, such as RedCap devices to support mid-market \gls{iot} use cases.
In this context, this work aims at reducing the energy consumption during beam management for RedCap devices, while ensuring that the desired Quality of Service (QoS) requirements are met. 
To do so, we formalize an optimization problem in an Indoor Factory (InF) scenario to select the best beam management parameters, including the beam update periodicity and the beamwidth, to minimize energy consumption based on users' distribution and their speed. 
The analysis yields the regions of feasibility, i.e., the upper limit(s) on the beam management parameters for RedCap devices, that we use to provide design guidelines accordingly.
\end{abstract}
	\begin{IEEEkeywords}
		5G NR, 3GPP, beam management, RedCap devices, energy consumption, Indoor Factory.
	\end{IEEEkeywords}
	
\begin{picture}(0,0)(0,-370)
\put(0,0){
\put(0,0){\qquad \qquad \quad This paper has been submitted to IEEE for publication. Copyright may change without notice.}}
\end{picture}	
	
\section{Introduction}
In the last few years, standardization bodies and industry players have developed several \gls{lpwan} technologies, such as \gls{lora}, \gls{nbiot}, and SigFox to support \gls{iot} applications in many fields, ranging from agriculture, transportation, logistics, and healthcare, as well as for smart cities~\cite{atzori2010internet,zanella2014internet}.
Along these lines, the \gls{3gpp} is also promoting new specifications~\cite{38875} to simplify 5G \gls{nr} standard operations to support high-end \gls{iot} devices, referred to as RedCap devices~\cite{varsier20215g}. 

Among other features, RedCap devices may be operating in the lower part of the \gls{mmwave} spectrum to improve the network performance in more demanding scenarios, such as in an indoor factory scenario~\cite{pagin2023nrlight}. 
Communication at \glspl{mmwave}, however, requires directionality between the transmitter and the receiver to compensate for the additional path loss experienced at those frequencies, typically realized via \gls{mimo} antenna arrays. In 5G \gls{nr}, beam management was designed to allow the endpoints to identify and  continuously maintain the optimal direction of 
transmission, e.g., during initial access and/or channel estimation~\cite{giordani2019tutorial}.
Specifically, beam management implies exhaustive search based on \glspl{ssb}, collected into bursts and transmitted by a \gls{gnb} according to pre-specified intervals and directions. 
However, beam management involves severe energy consumption for sending and receiving control signals, which is a function of the beamwidth and periodicity of \glspl{ssb}~\cite{giordani2017improved}.
Even though this is generally not an issue for 5G \gls{nr} systems, it may be challenging to handle for low-complexity, battery-powered RedCap devices, and may degrade the network performance.

Recently, the scientific community has explored possible simplifications of the 5G NR standard to optimize power consumption for RedCap devices~\cite{pagin2023nrlight}, for example via simplified air interface procedures~\cite{8613274}, protocol stack, antenna configurations~\cite{zhang2018low}, and enhanced power-saving functionalities such as Discontinuous Reception (eDRX) or wake-up signals~\cite{sultania2018energy}.
The \gls{3gpp} has also launched some Study and Work Items in this domain, for example in TR 38.869~\cite{38869} to study low-power wake-up signal and receiver for RedCap devices. 
However, to the best of our knowledge, there is no prior work focusing on beam management for RedCap devices, which stimulates further research in this domain.

To fill these gaps, in this work we formalize an optimization problem to determine the optimal beam management design for RedCap devices to minimize the energy consumption.
Notably, we focus on an \gls{inf} scenario, and derive the so-called regions of feasibility, i.e., the upper limit(s) on the beam management parameters, including the number of SSBs per burst and the burst periodicity, to guarantee that the \gls{qos} constraints are met, for example that \glspl{ue} never go undetected and/or maintain alignment as they move. 
Simulation results demonstrate that there exists an optimal configuration for beam management to promote energy efficiency, which depends on the speed of the \glspl{ue}, the beamwidth, and other network parameters.  


The rest of the paper is organized as follows. 
In Sec.~\ref{sec2} we present our system model (deployment, energy, mobility, and beam management). In Sec.~\ref{sec3} we describe our optimization problem. Also, we describe the impact of the number of antennas at the gNB on the QoS constraints. In Sec.~\ref{results} we present the simulation results and provide design guidelines towards the optimal set of parameters for beam management. Finally, conclusions are given in Sec.~\ref{conclude}.

\section{System Model}\label{sec2}
In this section we present our deployment model (Sec.~\ref{sub:deployment}), beam management model (Sec.~\ref{sub:beam-management}), energy consumption model (Sec.~\ref{sub:ec}), and mobility model (Sec.~\ref{sub:mob}).

\subsection{Deployment Model}
\label{sub:deployment}

We consider a 3GPP \gls{inf-sh} scenario~\cite{3GPPrel16} with an area of size $L\times W\times H$, a single \gls{gnb} placed at the center of the ceiling at height $h_{\rm gNB}$, and obstacles in the form of clutters of size $d_{c}$, height $h_{c}$, and density~$r$. 
Then, $K$ UEs are uniformly deployed at height $h_{\rm UE}$ around the clutters. 
The location of UE$_k$, for $k\in \{1, 2, \dots, K\}$, is given by $(d_k, \phi_k)$, where $d_k$ is the distance between UE$_k$ and the gNB, and $\phi_k$ is the phase of UE$_k$ measured counterclockwise. 
The UEs are assumed to be moving on a circle at constant velocity $v$ in a counterclockwise direction. 

The \gls{snr} $\gamma_k$ at UE$_k$ is given by~\cite{rawat2023optimal}
	\begin{equation}
		\gamma_{k}(d_{\rm3D})=
		\frac{\mathcal{H}_{\rm L}P_r(d_{\rm 3D})+\mathcal{H}_{\rm N}(1-P_r(d_{\rm 3D}))}{N_0\cdot B \cdot \text{NF}/G_{\text{gNB},k}G_{\rm UE}},
		\label{gamma_k}
	\end{equation}
where $d_{\rm 3D}=\sqrt{(h_{\rm gNB}-h_{\rm UE})^2+d_k^2}$ is the distance between the gNB and UE$_k$, $N_0$ is the noise Power Spectral Density, $B$ is the channel bandwidth, and NF is the noise figure. 
$\mathcal{H}_{\rm L}$ and $\mathcal{H}_{\rm N}$ include the effect of path loss, shadowing and fading parameter for the \gls{los} and \gls{nlos} channels, respectively and $P_r(d_{\rm 3D})$ is the \gls{los} probability, as described in~\cite{3GPPrel16}. Specifically, we have
\begin{equation}
	\mathcal{H}_{ j}=|\mathbbm{h}_{j}^{k}|^2 \text{PL}_{j}^{k},\: j\in\{\text{L,N}\},
\end{equation}
where $\mathbbm{h}_{j}^{k}$ and PL$_{j}^k$ are the channel fading gain and path loss for the \gls{los} (L) and \gls{nlos} (N) links, respectively.

In Eq.~\eqref{gamma_k}, $G_{\text{gNB},k}$ ($G_{\text{UE}}$) is the beamforming gain at the gNB (UE).
We assume analog beamforming (a realistic assumption for RedCap devices to minimize the energy consumption~\cite{8057288}), such that the gNB (UEs) can probe only one direction at a time. 
Specifically, the gNB is equipped with $N_{\rm gNB}$ antennas, and the beamforming gain is expressed as~\cite{balanis}
	\begin{align}
		G_{\text{gNB},k}=\sin\left(\dfrac{N_{\rm gNB}\pi}{2}\sin{\theta}_k\right)/\sin\left(\dfrac{\pi}{2}\sin{\theta}_k\right),
  \label{gain_gNB}
	\end{align}
where $\theta_{k}$ is the angular offset with respect to UE$_k$, as described in Sec.~\ref{sub:mob}.

\subsection{Beam Management Model}
\label{sub:beam-management}
According to the 5G \gls{nr} specifications~\cite{38300}, beam management operations rely on a directional version of the 3GPP LTE synchronization signal called \gls{ssb}.
Specifically, each SSB consists of 4 OFDM symbols in time and $240$ subcarriers in frequency, where the subcarrier spacing depends on the 5G NR numerology \cite{giordani2019tutorial}. 
Each SS block is mapped into a certain angular direction so that directional measurements can be made based on the quality of the received signal, e.g., in terms of the \gls{snr}.
To reduce the overhead,  SSBs can be gathered together into SS bursts.
An SS burst consists of $N_{\rm SS}\in \{8, 16, 32, 64\}$ SSBs, and the periodicity between consecutive SS bursts is $T_{\rm SS}\in \{5, 10, 20, 40, 80, 160\}$ ms. 

\subsection{Energy Consumption Model}
\label{sub:ec}

In 5G \gls{nr} beam management, the gNB transmits the SSBs by sequentially sweeping different angular directions to cover the whole beam space (or cell sector).
At the UE, the \gls{ec} required to receive those SSBs is equal to
\begin{equation}
	\text{EC}=S_{ D} P_{\rm UE}T_{\rm SSB},
	\label{EC1}
\end{equation}
where $S_{ D}$ is the number of SSBs required to completely sweep the beam space (which is a function of the beamwidth at the gNB), $P_{\rm UE}$ is the power consumed for receiving each SSB at the UE, and $T_{\rm SSB}$ is the time required to send each SSB.

From \cite[Eq. (3)]{giordani2019tutorial}, the number of SSBs required to completely sweep the beam space on the horizontal plane, with azimuth ranging from $0$ to $2\pi$, can be expressed~as 
\begin{align}
	S_D=\left\lceil{2\pi}/{\Delta_{3 \rm dB}}\right\rceil \approx \lceil\pi N_{\rm gNB}\rceil,
	\label{eq:SD}
\end{align}
where $\Delta_{3 \rm dB}$ is the 3-dB beamwidth, which can be approximated as $\Delta_{3 \rm dB}\approx 2/N_{\rm gNB}$ according to~\cite{balanis}.
Since each SSB consists of 4 OFDM symbols, the time required to send one SSB can be expressed as \cite[Eq. (2)]{giordani2019tutorial}
\begin{align}
	T_{\rm SSB}=4T_{\rm symb}=4\left({71.45}/{2^n}\right),
\end{align}
where $n$ represents the 5G \gls{nr} numerology.

Finally, the power consumption at the UE, equipped with $N_{\rm UE}$ antennas, can be expressed as
\begin{equation}
	P_{\rm UE}=N_{\rm UE}(P_{\rm LNA}+P_{\rm PS})+P_{\rm RF}+P_C+2P_{\rm ADC},
	\label{eq:power}
\end{equation}
where $P_{\rm RF}=P_M+P_{\rm LO}+P_{\rm LPF}+P_{\rm BB}$ is the power consumption of the RF chain~\cite{mendez2016}. 
A description of the power components appearing in Eq.~\eqref{eq:power}, and the relative numerical values, is provided in Table~\ref{parameter}.


\begin{figure}
	\centering  
 \setlength\belowcaptionskip{-0.8cm}
	\includegraphics[width=.65\columnwidth]{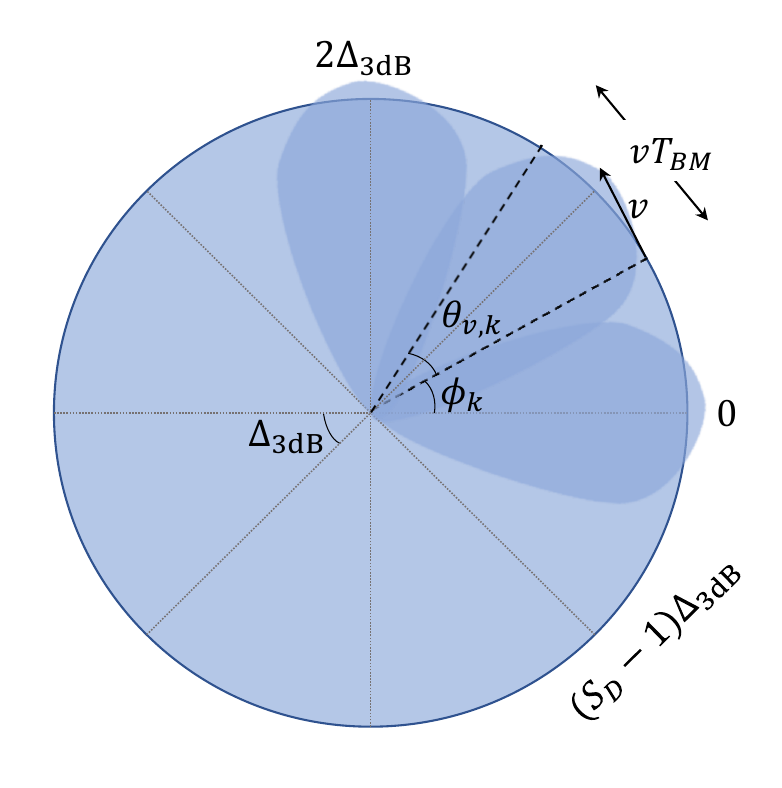}
	\caption{UE mobility model. During beam management, UE$_k$ accumulates an angular offset ${\theta}_k$ due to both initial misalignment ($\theta_{i,k}$) and mobility ($\theta_{v,k}$) The latter depends on the \gls{ue} speed $v$, and the beam management time $T_{\rm BM}$.\vspace{-1cm}}
	\label{vel_model}
\end{figure}

\subsection{Mobility Model}
\label{sub:mob}
At the beginning of the beam management process, UE$_k$, $k\in \{1, 2, \dots, K\}$, establishes a physical link connection with the gNB using a certain beam.
Due to the finite pre-defined codebook of directions available at the gNB,
UE$_k$ comes with a non-zero initial angular offset $\theta_{i,k}=\min(\phi_k-\bm{\bar{\phi}_{br}})$ with respect to the gNB antenna boresight directions $\bm{\bar{\phi}_{br}} \doteq \left[ 0, 1, 2, \dots, S_D-1 \right] \Delta_{3\rm dB}$, 
as represented in Fig.~\ref{vel_model}.

At the same time we assume that, during beam management, UE$_k$ can move in a counterclockwise direction at constant velocity $v$.
During this time, UE$_k$ may lose beam alignment and the corresponding beamforming gain, and get disconnected if the resulting SNR is lower than a pre-defined threshold~\cite{giordani2018coverage}. 
Thus, we define $\theta_{v,k}$ as the angular offset due to mobility during beam management, i.e.,
\begin{equation}
	\theta_{v,k}={v T_{\rm BM}}/{d_k}.
	\label{eq:phi}
\end{equation}
In Eq.~\eqref{eq:phi}, $T_{\rm BM}$ is the time for beam management, and is measured as the delay from the first SSB transmission to the completion of the sweep in all possible angular directions, which can be expressed as in~\cite[Eq.~(4)]{giordani2019tutorial}, i.e.,
\begin{align}
	T_{\rm BM}=T_{\rm SS}\left(\left\lceil{S_D}/{N_{\rm SS}}\right\rceil-1\right)+T_{\ell},
	\label{T_BM}
\end{align}
where $T_{\ell}$ is the time required to send the remaining SSBs in the last burst and is given in \cite[Eq.~(6)]{giordani2019tutorial}. 
%
%
 Therefore, the overall angular offset for UE$_k$ during beam management, due to both the initial offset ($\theta_{i,k}$) and the offset accumulated due to mobility ($\theta_{v,k}$), can be expressed as 
\begin{equation}
	\theta_k=|\theta_{v,k}+\theta_{i,k}|.
	\label{theta_k}
\end{equation}
\section{Optimization Problem}\label{sec3}
In this section we define an optimization problem to minimize the energy consumption for RedCap devices for transmitting/receiving SSBs during beam management. 
The optimization problem can be formalized as follows:
\begin{subequations}
	\begin{alignat}{2}
		&\!\min_{N_{\rm gNB}} \quad       &\text{EC}=&S_D P_{\rm UE}T_{\rm SSB}, \label{EC2}\\
		& & & P_T\gamma_k\geq \tau, \: \forall k;\label{C1}\\
		& & & N_{\rm gNB}\in \{2,3, \dots, 64\},\label{C2}
	\end{alignat}
\label{OP}
\end{subequations}
where $P_T$ is the transmission power at the gNB, and $\gamma_k$ is the SNR at UE$_k$ as given in Eq.~\eqref{gamma_k}.
In~\eqref{OP},~\eqref{C1} ensures that the SNR at UE$_k$ is greater than or equal to a minimum threshold $\tau$, which is large enough to ensure that UE$_k$ can be properly detected, and~\eqref{C2} restricts the number of antenna elements at the gNB to $64$, as expected for RedCap~devices.

\textbf{Modeling of the constraints.} 
The optimization problem determines the optimal value of $N_{\text{gNB}}$, referred to as $N^*$, based on the SNR $\gamma_k$, $\forall k$, which depends on $G_{\rm gNB}$, so on the angular offset ${\theta}_k$ introduced by the moving UEs. 
Indeed, as the UE moves at constant velocity $v$ during the beam management process, it may lose alignment with respect to the associated beam, potentially deteriorating the beamforming gain. This may cause the SNR of UE$_k$ to drop below the sensitivity threshold $\tau$, preventing it from being detected. The factors that may lead to misalignment include:
(i) the UE velocity $v$ (the faster the UE, the sooner it may lose alignment); (ii) the beam management time $T_{\rm BM}$ and, consequently $T_{\rm SS}$ and $N_{\rm SS}$ (the slower the beam management procedure, the higher the probability that the UE would lose alignment); and (iii) the number of antennas $N_{\rm gNB}$, which defines the beamwidth (the narrower the beam, the higher the probability that the UE would lose alignment).
In the following, we investigate the impact of those terms on the optimization problem.

\begin{figure}
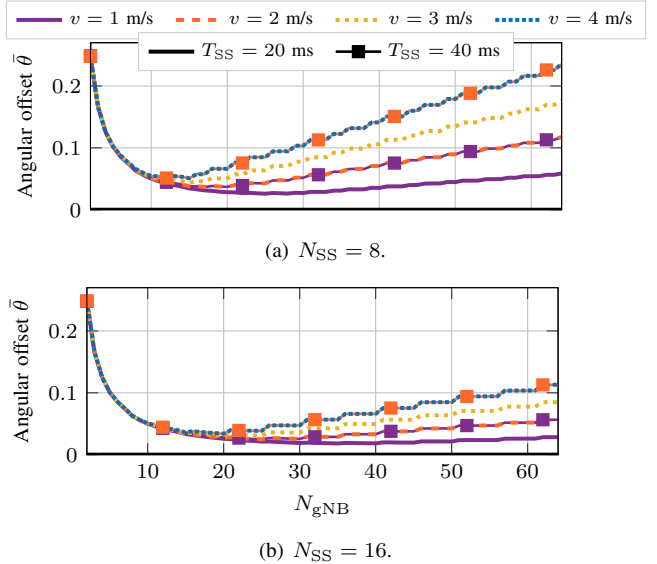

	\centering 
    \setlength\fwidth{0.7\columnwidth}
    \setlength\fheight{0.25\columnwidth}
	\subfigure[$N_{\rm SS}=8$.] {\input{Figures/theta_avg_Nss8}\label{thetahat_avg_N_Nss8}} 
    \centering 
    \setlength\fwidth{0.7\columnwidth}
    \setlength\fheight{0.25\columnwidth}
	\subfigure[$N_{\rm SS}=16$.] {\hspace{-1.2cm}\input{Figures/theta_avg_Nss16}\label{thetahat_avg_N_Nss16}}
	\caption{Average angular offset $\bar{\theta}$, as a function of $N_{\rm gNB}$, the \gls{ue} speed $v$, and the SS burst periodicity $T_{\rm SS}$.\vspace{-2cm}}\label{thetahat_avg}
\end{figure}
\begin{figure}
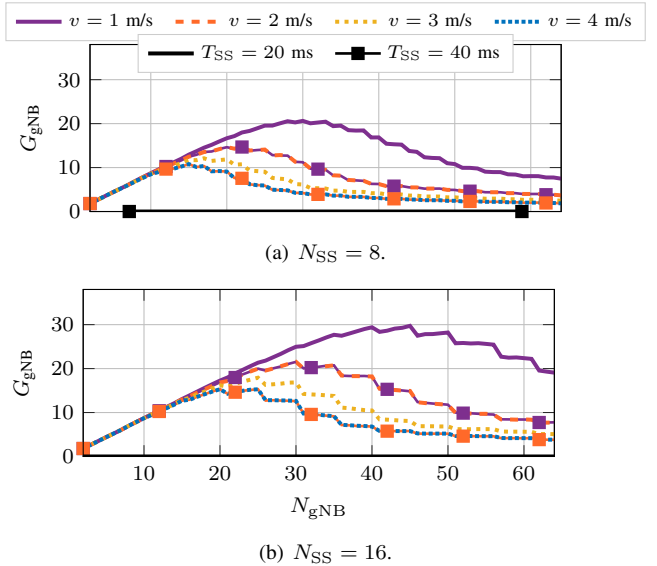

	\centering  
    \setlength\fwidth{0.7\columnwidth}
    \setlength\fheight{0.25\columnwidth}
	\subfigure[$N_{\rm SS}=8$.] {\input{Figures/Gain_avg_Nss8}\label{Gain_avg_N_Nss8}} 
    \setlength\fwidth{0.7\columnwidth}
    \setlength\fheight{0.25\columnwidth}
	\subfigure[$N_{\rm SS}=16$.] {\hspace{-1.2cm}\input{Figures/Gain_avg_Nss16}\label{Gain_avg_N_Nss16}}
	\caption{Average gain at the gNB ${G}_{\textrm{gNB}}$ as a function of $N$, the \gls{ue} speed $v$, and the SS burst periodicity $T_{\rm SS}$.\vspace{-2cm}}\label{Gain_avg_N_Nss}
\end{figure}

In Fig. \ref{thetahat_avg} we plot $\bar{\theta}$ (the angular offset averaged over all $K$ UEs in the scenario) vs. $N_{\rm gNB}$ for different values of $v$ and $T_{\rm SS}$, and for $N_{\rm SS}=\{8, 16\}$. 
We observe that $\bar{\theta}$ initially decreases with $N_{\rm gNB}$. In fact, when the number of antennas is small, the beam is large enough to ensure continuous alignment despite mobility. In this region, $\bar{\theta}$ is thus dominated by the initial offset $\theta_{i,k}$ with respect to the antenna boresight direction. 
Then, as $N_{\rm gNB}$ increases, the beams become progressively narrower, and the number of SSBs that are required to be sent to completely sweep the beam space also increases, which increases the beam management time. 
In these conditions, the angular offset due to mobility ${\theta}_{v,k}$ increases accordingly as per Eq.~\eqref{eq:phi}.
In addition, we observe that in both Fig. \ref{thetahat_avg_N_Nss8} and \ref{thetahat_avg_N_Nss16} the angular offset for $v=2$ m/s and $T_{\rm SS}=20$ ms overlaps with the offset for $v=1$ m/s and $T_{\rm SS}=40$ ms. Similarly, the offset for $v=2$ m/s and $T_{\rm SS}=40$ overlaps over the offset for $v=4$ m/s and $T_{\rm SS}=20$ ms. Therefore, we conclude that the angular offset depends on $v$ and $T_{\rm SS}$ only through their product. This observation becomes significant while analyzing the feasibility regions in Sec.~\ref{results}.

Notice that the zigzag effect in Fig.~\ref{thetahat_avg} is due to the fact that $\theta_{v,k}$ and hence $\bar{\theta}$ is a function of $T_{\rm BM}$ which, as reported in Eq.~\eqref{T_BM}, is a ceiling function. This effect increases as $N_{\rm gNB}$ increases, i.e., as $\theta_{v,k}$ dominates the average angular offset $\bar{\theta}$.


Additionally, in Fig.~\ref{Gain_avg_N_Nss} we plot the average antenna gain at the gNB ($G_{\rm gNB}$, averaged over all $K$ UEs in the scenario) vs. $N_{\rm gNB}$ for different values of $v$ and $T_{\rm SS}$, and for $N_{\rm SS}=\{8, 16\}$. 
We notice that $G_{\rm gNB}$ initially increases with $N_{\rm gNB}$, and then drops after a threshold due to mobility.
If $N_{\rm M}$ is the number of antennas corresponding to the point where $G_{\rm gNB}$ is maximum, we conclude that the optimization problem in~\eqref{OP} is restricted to the values of $N_{\rm gNB}\leq N_{\rm M}$ because the energy consumption increases with $N_{\rm gNB}$. 
We further observe that $N_{\rm M}$ decreases with $v$ and $T_{\rm SS}$, and increases with $N_{\rm SS}$.
In other words, the product $v T_{\rm SS}$ for a given value of $N_{\rm SS}$ establishes an upper limit to determine the regions of feasibility, as further discussed in Sec.~\ref{results}: if the SNR constraints are not satisfied for $N_{\rm gNB}\leq N_{\rm M}$, the optimization problem will be infeasible. 

\textbf{Optimization algorithm.}
\label{sec:alg}
Based on the optimization problem in~\eqref{OP}, and the considerations above, we conclude that the energy consumption at the UE increases monotonically with the number of antennas at the gNB. This suggests that the minimum value of $N_{\rm gNB}$ at which the SNR constraints are satisfied should be the optimal $N_{\rm gNB}$, or $N^*$. 
For a given transmit power ($P_T$) and SNR threshold ($\tau$), if the constraints are not met, the problem is infeasible.



\section{Numerical Results}\label{results}
In this section, we evaluate the energy consumption for beam management as a function of $N_{\rm SS}$, $T_{\rm SS}$, $N_{\rm gNB}$, $v$, $P_T$, and $\tau$.
Specifically, we perform $10^5$ Monte Carlo simulations in MATLAB for each combination of parameters, and in each simulation we find $N^*$ using the algorithm presented in Sec.~\ref{sec:alg}. The simulation parameters are reported in Tab.~\ref{parameter}, taken from~\cite[Table 7.2-4]{3GPPrel16} for the \gls{inf-sh} scenario, \cite{pagin2023nrlight} for the RedCap devices, and~\cite{mendez2016} for the power consumption.  

The goal of our analysis is to determine the regions of feasibility, and the corresponding set of 5G \gls{nr} beam management parameters which minimize the energy consumption while satisfying SNR constraints. 
Notice that we assume zero misdetection probability in the analysis, i.e., no user goes misdetected during the beam management process.

\begin{table}[t!]
\centering
\scriptsize
\renewcommand{\arraystretch}{1.1}
\caption{Simulation parameters.}
\label{parameter}
	\begin{tabular}{ l|l|l }
		\hline
		{Parameter} & {Description} & {Value}\\
		\hline
		$h_{\rm gNB}$&  gNB height  & $25$ m\\
		$h_{\rm UE}$ & UE height & $1.5$ m\\
		$d_c$    & Clutter size & $10$ m \\
		$h_c$&  Clutter height & $5$ m\\
		$r$ & Clutter density & $20\%$ \\
		$L\times W\times H$ &  Size of the \gls{inf-sh} scenario & $20\times20\times25$ m \\ \hline
		$K$   & Number of UEs  & $50$\\
        $f_c$   & Carrier frequency  & $28$ GHz\\
        $B$ & Bandwidth  & $50$ MHz\\
        $P_T$ & Transmission power  & $18$ dBm\\
        $\tau$ & SNR threshold  & $\{3, 7, 10\}$ dB\\
		$\mathbbm{h}_{\rm LoS}^{k}, \mathbbm{h}_{\rm NLoS}^{k}$ & Channel fading gains & $\mathcal{CN}(0,1)$ \\
        $\text{PL}_{\rm LoS}^k$, $\text{PL}_{\rm NLoS}^k$ & Path loss & \cite[Table 7.4.1-1]{3GPPrel16}\\
        $P_r(d_{3D})$ & LoS probability & \cite[Table 7.4.2-1]{3GPPrel16} \\
        $N_0$ & Noise Power Spectral Density  & $-174$ dBm/Hz\\
        NF & Noise figure  & $9$ dB\\
        $n$ & 5G NR numerology index & $4$ \\\hline
        $P_{\rm LNA}$   & Low noise amplifier  power & $20$ mW\\
 		$P_{\rm PS}$&  Phase shifter  power & $30$ mW\\
 		$P_M$ & Mixer power & $19$ mW\\
 		$P_{\rm LO}$    &Local oscillator power &  $5$ mW\\
 		$P_{\rm LPF}$&  Low pass filter power & $14$ mW\\
 		$P_{\rm BB}$& Baseband amplifier power & $5$ mW\\
 		$P_{\rm ADC}$& ADC  power & $200$ mW\\
		\hline
	\end{tabular}
\end{table}
\vspace{-0.5em}

\subsection{Impact of $T_{\rm SS}$ and $N_{\rm SS}$}\label{results1}
\begin{figure}
\centering  
    \setlength\fwidth{0.7\columnwidth}
    \setlength\fheight{0.35\columnwidth}
    \subfigure[Misdetection probability.]{
%
%
\definecolor{mycolor1}{rgb}{0.00000,0.44700,0.74100}%
\definecolor{mycolor2}{rgb}{0.85000,0.32500,0.09800}%
\definecolor{mycolor3}{rgb}{0.92900,0.69400,0.12500}%
\definecolor{mycolor4}{rgb}{0.46600,0.67400,0.18800}%
\definecolor{mycolor5}{rgb}{0.49412,0.18431,0.55686}%
\begin{tikzpicture}

\pgfplotsset{every tick label/.append style={font=\footnotesize}}

 \begin{axis}[
    width=\fwidth,
    height=\fheight,
    at={(0\fwidth,0\fheight)},
    scale only axis,
    legend image post style={mark indices={}},
    legend style={
        /tikz/every even column/.append style={column sep=0.2cm},
        at={(0.5, 1.02)}, 
        anchor=south, 
        draw=white!80!black, 
        font=\scriptsize
        },
    legend columns=3,
    xlabel style={font=\footnotesize},
    xlabel={},
    xmajorgrids,
    xtick={5, 10, 20, 40, 80, 160},
    xmin=2,
    xmax=160,
    xtick style={color=white!15!black},
    ylabel shift = -1 pt,
    ylabel style={font=\footnotesize},
    ylabel={Misdetection probability},
    ymajorgrids,
    ymajorticks=true,
    ymin=0,
    ymax=0.58,
    ytick style={color=white!15!black}
]

\addplot [color=mycolor1, line width=1.5pt, mark=diamond, mark options={solid, mycolor1}]
  table[row sep=crcr]{%
5	0\\
10	0\\
20	0\\
40	0\\
80	0\\
160	0.0057\\
};
\addlegendentry{$v=1$ m/s}

\addplot [color=mycolor2, line width=2.0pt, mark=*, mark options={solid, mycolor2}]
  table[row sep=crcr]{%
5	0\\
10	0\\
20	0\\
40	0\\
80	0.007\\
160	0.1638\\
};
\addlegendentry{$v=2$ m/s}

\addplot [color=mycolor3, line width=1.5pt, mark=square*, mark options={solid, mycolor3}]
  table[row sep=crcr]{%
5	0\\
10	0\\
20	0\\
40	0.001\\
80	0.056\\
160	0.3226\\
};
\addlegendentry{$v=3$ m/s}

\addplot [color=mycolor4, dashed, line width=1.5pt]
  table[row sep=crcr]{%
5	0\\
10	0\\
20	0\\
40	0.0048\\
80	0.1734\\
160	0.456\\
};
\addlegendentry{$v=4$ m/s}

\addplot [color=mycolor5, dotted, line width=1.5pt, mark=o, mark options={solid, mycolor5}]
  table[row sep=crcr]{%
5	0\\
10	0\\
20	0.0004\\
40	0.023\\
80	0.2412\\
160	0.532\\
};
\addlegendentry{$v=5$ m/s}

\end{axis}

\begin{axis}[%
width=0.36\fwidth,
height=0.22\fwidth,
at={(0.088\fwidth,0.218\fwidth)},
scale only axis,
xmin=5.97571329051635,
xmax=100.424759812499,
xtick={5,20,80,160},
xlabel style={font=\color{white!15!black}},
xlabel={},
ymin=0,
ymax=0.0049,
ytick={0,0.002,0.004},
axis background/.style={fill=white},
xmajorgrids,
ymajorgrids
]
\addplot [color=mycolor1, line width=1.5pt, mark=diamond, mark options={solid, mycolor1}, forget plot]
  table[row sep=crcr]{%
5	0\\
10	0\\
20	0\\
40	0\\
80	0\\
160	0.0057\\
};
\addplot [color=mycolor2, line width=1.5pt, mark=*, mark options={solid, mycolor2}, forget plot]
  table[row sep=crcr]{%
5	0\\
10	0\\
20	0\\
40	0\\
80	0.007\\
160	0.1638\\
};
\addplot [color=mycolor3, line width=1.5pt, mark=square*, mark options={solid, mycolor3}, forget plot]
  table[row sep=crcr]{%
5	0\\
10	0\\
20	0\\
40	0.001\\
80	0.056\\
160	0.3226\\
};

\addplot [color=mycolor4, dashed, line width=1.5pt, forget plot]
  table[row sep=crcr]{%
5	0\\
10	0\\
20	0\\
40	0.0048\\
80	0.1734\\
160	0.456\\
};

\addplot [color=mycolor5, dotted, line width=1.5pt, mark=o, mark options={solid, mycolor5}, forget plot]
  table[row sep=crcr]{%
5	0\\
10	0\\
20	0.0004\\
40	0.023\\
80	0.2412\\
160	0.532\\
};
\end{axis}

\end{tikzpicture}%
\label{prob_md_Nss8}}
    \setlength\fwidth{0.7\columnwidth}
    \setlength\fheight{0.25\columnwidth}
    \subfigure[Average optimal number of antennas at the gNB ($N^*$).] {
%
%
\definecolor{mycolor1}{rgb}{0.00000,0.44700,0.74100}%
\definecolor{mycolor2}{rgb}{1.00000,0.41176,0.16078}%
\definecolor{mycolor3}{rgb}{0.92900,0.69400,0.12500}%
\definecolor{mycolor4}{rgb}{0.46667,0.67451,0.18824}%
\definecolor{mycolor5}{rgb}{0.49412,0.18431,0.55686}%

\begin{tikzpicture}

\pgfplotsset{every tick label/.append style={font=\footnotesize}}

 \begin{axis}[
    width=\fwidth,
    height=\fheight,
    at={(0\fwidth,0\fheight)},
    scale only axis,
    legend image post style={mark indices={}},
    legend style={
        /tikz/every even column/.append style={column sep=0.2cm},
        at={(0.5, 0.97)}, 
        anchor=south, 
        draw=white!80!black, 
        font=\scriptsize
        },
    legend columns=3,
    scale only axis,
    bar shift auto,
    xlabel style={font=\footnotesize},
    xlabel={SS Burst period $T_{\rm SS}$ (ms)},
    xmajorgrids,
    xmin=0.5, xmax=6.4,
    xtick style={color=white!15!black},
    xtick={1,2,3,4,5,6},
    xticklabels={{5},{10},{20},{40},{80},{160}},
    ylabel shift = -1 pt,
    ylabel style={font=\footnotesize},
    ylabel={$N^*$},
    ymajorgrids,
    ymajorticks=true,
    ymin=0,
    ymax=6.0,
    ytick style={color=white!15!black}
]

\addplot[ybar, bar width=0.115, preaction={fill, mycolor1}, pattern={Dots[distance=1mm, radius=0.25mm]}, pattern color=white, draw=black, area legend] table[row sep=crcr] {%
1	5.35005\\
2	5.34934\\
3	5.36382\\
4	5.38138\\
5	5.39409\\
};
\addlegendentry{$v=1$ m/s}

\addplot[ybar, bar width=0.115, preaction={fill, mycolor2}, pattern={horizontal lines}, pattern color=black, draw=black, area legend] table[row sep=crcr] {%
1	5.35239\\
2	5.35795\\
3	5.36284\\
4	5.39319\\
5	0\\
};
\addlegendentry{$v=2$ m/s}

\addplot[ybar, bar width=0.115, preaction={fill, mycolor3}, draw=black, pattern={north east lines}, pattern color=black, area legend] table[row sep=crcr] {%
1	5.36212\\
2	5.363\\
3	5.37313\\
4	0\\
5	0\\
};
\addlegendentry{$v=3$ m/s}

\addplot[ybar, bar width=0.115, preaction={fill, mycolor4}, draw=black, area legend] table[row sep=crcr] {%
1	5.35956\\
2	5.36651\\
3	5.38857\\
4	0\\
5	0\\
};
\addplot[forget plot, color=white!15!black] table[row sep=crcr] {%
0.507692307692308	0\\
5.49230769230769	0\\
};
\addlegendentry{$v=4$ m/s}

\addplot[ybar, bar width=0.115, preaction={fill, mycolor5}, draw=black, pattern={grid}, pattern color=white, area legend] table[row sep=crcr] {%
1	5.35761\\
2	5.37246\\
3	0\\
4	0\\
5	0\\
};
\addlegendentry{$v=5$ m/s}

\end{axis}
\end{tikzpicture}%
\label{Nopt_Tss}}
\caption{Misdetection probability (top) and $N^*$ (bottom) vs. the SS burst periodicity $T_{\rm SS}$ and the \gls{ue} speed $v$, considering $P_T=18$ dBm, sensitivity threshold $\tau=7$, $N_{\rm SS}=8$.\vspace{-1cm}}\label{SAR_R10}
\end{figure}
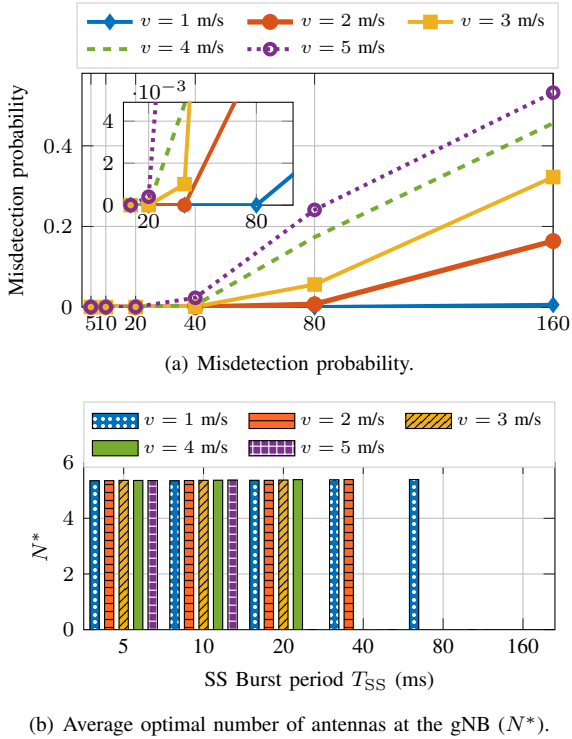

Figs.~\ref{prob_md_Nss8} and~\ref{Nopt_Tss} depict the UE misdetection probability and $N^*$, respectively, as a function of $T_{\rm SS}$ and $v$, for $N_{\rm SS}=8$. 
While $N^*$ depends on $P_T$ and $\tau$, it does not change with $v$, $T_{\rm SS}$, and $N_{\rm SS}$. 
This is because the objective function always drives the optimization problem towards the minimum value of $N_{\rm gNB}$ that meets the SNR constraints for each UE, so as to minimize the energy consumption. In turn, this sets $N^*$ to the minimum value corresponding to the largest angular offset beyond which the problem becomes infeasible (which determines the values of $v$ and $T_{\rm SS}$ for a given $N_{\rm SS}$), as described in Sec.~\ref{sec3}.

Nevertheless, given $P_T$ and $\tau$, there exists only a limited set of values of $v$, $T_{\rm SS}$ at $N_{\rm SS}$ 
for which the SNR constraints are met. As a consequence, some bars are missing in Fig.~\ref{Nopt_Tss}, which indicates that the corresponding problem is  infeasible. 
For example, for $T_{\rm SS}=160$ ms and $v\geq 1$ m/s, there are no values of $N_{\rm gNB}$ for which $\text{SNR}_k\geq \tau,~\forall k$. This is also observed in Fig.~\ref{prob_md_Nss8}, where the misdetection probability at $T_{\rm SS}=160$ ms is greater than zero for $v\geq 1$ m/s. 
Similarly, $v\geq 2$ m/s is infeasible for $T_{\rm SS}\geq80$ ms, whereas for $v\leq 4$ m/s and  $T_{\rm SS}\leq20$ ms the problem is feasible, which yields $N^*=5.4$ on average.\footnote{Notice that, while we constrain $N^*$ to be an integer in each Monte Carlo simulation, here $N^*$ represents the average of different realizations.} 
This is because increasing $v$ or $T_{\rm SS}$ increases the average angular offset as per Eq.~\eqref{theta_k}, and may cause the UEs to lose beam alignment sooner, thus making the problem infeasible. 


\subsection{Impact of $P_T$ and $\tau$}
\begin{figure}
\centering  
    \setlength\fwidth{0.75\columnwidth}
    \setlength\fheight{0.25\columnwidth}
    \subfigure[$\tau=3$ dB.]{\hspace{-0.4cm}\definecolor{mycolor1}{rgb}{0.00000,0.44700,0.74100}%
\definecolor{mycolor2}{rgb}{1.00000,0.41176,0.16078}%
\definecolor{mycolor3}{rgb}{0.92900,0.69400,0.12500}%
\definecolor{mycolor4}{rgb}{0.46667,0.67451,0.18824}%

\pgfplotsset{every tick label/.append style={font=\footnotesize}}

\begin{tikzpicture}

\begin{axis}[
    width=\fwidth,
    height=\fheight,
    at={(0\fwidth,0\fheight)},
    scale only axis,
    legend image post style={mark indices={}},
legend style={
        /tikz/every even column/.append style={column sep=0.2cm},
        at={(0.5, 0.9)}, 
        anchor=south, 
        draw=white!80!black, 
        font=\scriptsize
        },
    legend columns=2,
    scale only axis,
    bar shift auto,
    xlabel style={font=\footnotesize},
    xlabel={},
    xmajorgrids,
    xmajorticks=false,
    xmin=11.5, xmax=18.5,
    xtick style={color=white!15!black},
    xtick={12,13,14,15,16,17,18},
    ylabel style={font=\footnotesize},
    ylabel={$N^*$},
    ymin=0, ymax=9,
    ymajorgrids,
    ymajorticks=true,
    ytick style={color=white!15!black}
]

\addplot[ybar, bar width=0.165, preaction={fill, mycolor1}, pattern={Dots[distance=1mm, radius=0.25mm]}, pattern color=white, draw=black, area legend] table[row sep=crcr] {%
12	7.4262\\
13	6.3148\\
14	5.393\\
15	4.5976\\
16	3.884\\
17	3.4016\\
18	3.027\\
};
\addlegendentry{$T_{ss}=20$ m/s}

\addplot[ybar, bar width=0.165, preaction={fill, mycolor2}, pattern={horizontal lines}, pattern color=black, draw=black, area legend] table[row sep=crcr] {%
12	7.4926\\
13	6.3544\\
14	5.3834\\
15	4.5642\\
16	3.9036\\
17	3.4068\\
18	3.0048\\
};
\addlegendentry{$T_{ss}=40$ m/s}

\addplot[ybar, bar width=0.165, preaction={fill, mycolor3}, draw=black, pattern={north east lines}, pattern color=black, area legend] table[row sep=crcr] {%
12	0\\
13	0\\
14	5.3778\\
15	4.594\\
16	3.9668\\
17	3.402\\
18	3.0302\\
};
\addlegendentry{$T_{ss}=80$ m/s}

\addplot[ybar, bar width=0.165, fill=mycolor4, draw=black, area legend] table[row sep=crcr] {%
12	0\\
13	0\\
14	0\\
15	0\\
16	0\\
17	3.4118\\
18	3.0184\\
};
\addlegendentry{$T_{ss}=160$ ms}

\end{axis}
\end{tikzpicture}\label{P_md_Nss8}}{\hspace{-0.4cm}}
    \setlength\fwidth{0.75\columnwidth}
    \setlength\fheight{0.25\columnwidth}
    \subfigure[$\tau=7$ dB.]{\hspace{-0.4cm}\definecolor{mycolor1}{rgb}{0.00000,0.44700,0.74100}%
\definecolor{mycolor2}{rgb}{1.00000,0.41176,0.16078}%
\definecolor{mycolor3}{rgb}{0.92900,0.69400,0.12500}%
\definecolor{mycolor4}{rgb}{0.46667,0.67451,0.18824}%

\pgfplotsset{every tick label/.append style={font=\footnotesize}}

\begin{tikzpicture}

\begin{axis}[
    width=\fwidth,
    height=\fheight,
    at={(0\fwidth,0\fheight)},
    scale only axis,
    legend image post style={mark indices={}},
legend style={
        /tikz/every even column/.append style={column sep=0.2cm},
        at={(0.5, 0.97)}, 
        anchor=south, 
        draw=white!80!black, 
        font=\scriptsize
        },
    legend columns=2,
    scale only axis,
    bar shift auto,
    xlabel style={font=\footnotesize},
    xlabel={$P_T$ (dBm)},
    xmajorgrids,
    xmin=11.5, xmax=18.5,
    xtick style={color=white!15!black},
    xtick={12,13,14,15,16,17,18},
    ylabel style={font=\footnotesize},
    ylabel={$N^*$},
    ymin=0, ymax=13,
    ymajorgrids,
    ymajorticks=true,
    ytick style={color=white!15!black}
]

\addplot[ybar, bar width=0.165, preaction={fill, mycolor1}, pattern={Dots[distance=1mm, radius=0.25mm]}, pattern color=white, draw=black, area legend] table[row sep=crcr] {%
12  0\\
13  12.368\\
14  10.4658\\
15  8.7942\\
16  7.4488\\
17  6.2616\\
18  5.3698\\
};

\addplot[ybar, bar width=0.165, preaction={fill, mycolor2}, pattern={horizontal lines}, pattern color=black, draw=black, area legend] table[row sep=crcr] {%
12  0\\
13  0\\
14  0\\
15  8.8956\\
16  7.52\\
17  6.3312\\
18  5.4008\\
};

\addplot[ybar, bar width=0.165, preaction={fill, mycolor3}, draw=black, pattern={north east lines}, pattern color=black, area legend] table[row sep=crcr] {%
12  0\\
13  0\\
14  0\\
15  0\\
16  0\\
17  6.3886\\
18  5.397\\
};

\end{axis}
\end{tikzpicture}\label{P_Tss}}
	\caption{Optimal number of antennas at the gNB $N^*$ vs. $P_T$ and SS burst periodicity $T_{\rm SS}$, for sensitivity threshold $\tau \in \{3, 7 \}$~dB, $N_{\rm SS}=8$, and \gls{ue} speed $v=1$ m/s.\vspace{-2cm}}
    \label{P}
\end{figure}
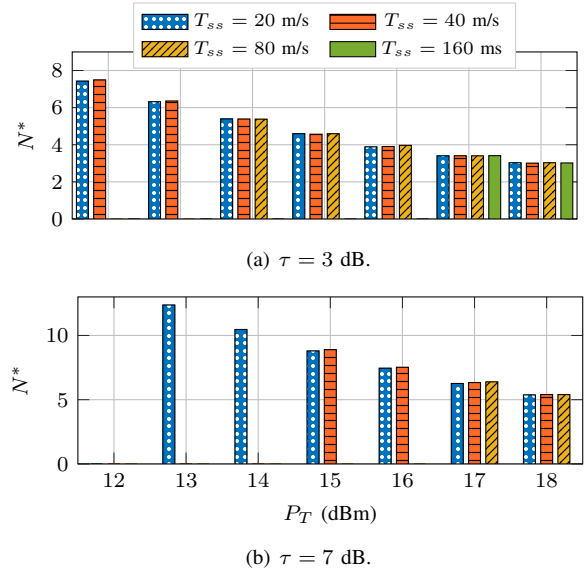

Fig.~\ref{P} illustrates the average optimal number of antennas $N^*$ as a function of the transmission power $P_T$ and $T_{\rm SS}$, for $\tau\in\{3, 7\}$ dB, $v=1$ m/s and $N_{\rm SS}=8$. We observe that as $P_T$ decreases and $\tau$ increases, $N^*$ increases. 
Indeed, increasing the number of antennas leads to a higher (best case) beamforming gain, thus possibly improving the minimum SNR at the UEs. 
At the same time, decreasing $P_T$ and/or increasing~$\tau$ also reduces the set of values for which the problem is feasible, as demonstrated by the missing bars in Fig.~\ref{P}. In fact, although the angular offset does not directly depend on $P_T$ and $\tau$, a smaller $P_T$ or a higher $\tau$ effectively impose progressively stricter constraints on the problem, as per C1 in~\eqref{OP}. For instance, \{$P_T=18$~dBm, $\tau = 3$~dBm, $T_{\rm SS}\leq 160$~ms\} is a feasible configuration, whereas \{$P_T=12$~dBm, $\tau = 3$~dBm, $T_{\rm SS}> 40$ ms\} is not. 

\subsection{Feasibility Regions}
For given values of $P_T$ and $\tau$, there exists only a limited set of values of $v$, $T_{\rm SS}$, and $N_{\rm SS}$ for which the problem in \eqref{OP} is feasible, i.e., the SNR constraints are guaranteed. 
Table~\ref{FR} reports these feasibility regions for $P_T=18$~dBm and $\tau\in\{ 3,7, 10 \}$~dB, in terms of the highest product of $v$ and $T_{\rm SS}$ supported by the system. 
We recall that, as observed in Sec.~\ref{sec3}, both $v$ and $T_{\rm SS}$ have the same impact on the angular offset, and the feasibility regions are perfectly defined by the product $vT_{SS}$.
For example, for $N_{\rm SS}=8$ and $\tau=7$~dB, the feasibility region is upper bounded by $vT_{SS}=0.08$ m. The results in Table~\ref{FR} have been obtained for different values of $N_{\rm SS}$ and $\tau$, and $v\leq 25$ m/s using a similar analysis as in Sec.~\ref{results1}.



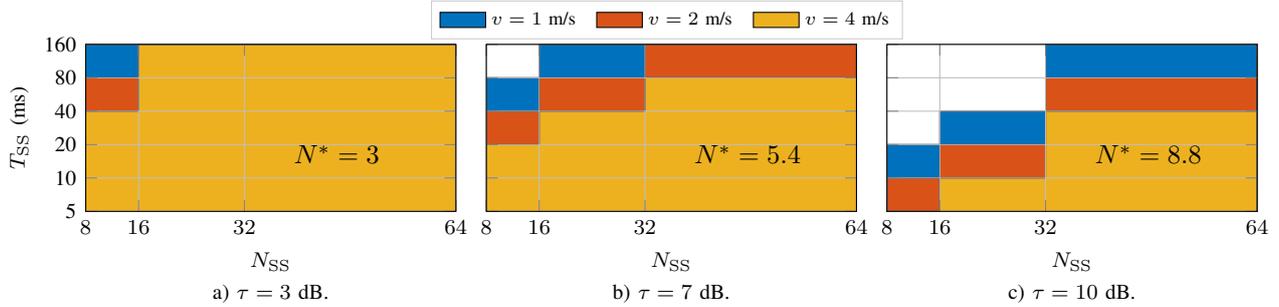
\begin{figure*}
	\centering  
    \setlength\belowcaptionskip{-0.99cm}
    \setlength\fwidth{0.55\columnwidth}
    \setlength\fheight{0.25\columnwidth}
	\begin{tikzpicture}

\definecolor{mycolor1}{rgb}{0.00000,0.44700,0.74100}%
\definecolor{mycolor2}{rgb}{0.85000,0.32500,0.09800}%
\definecolor{mycolor3}{rgb}{0.92900,0.69400,0.12500}%

\pgfplotsset{every tick label/.append style={font=\footnotesize}}

\begin{groupplot}[
  group style={
    group size=3 by 1,
    group name=plots,
    horizontal sep= 0.4 cm
  },
    width=\fwidth,
    height=\fheight,
    at={(0\fwidth,0\fheight)},
    scale only axis,
    legend image post style={mark indices={}},
    legend style={
        /tikz/every even column/.append style={column sep=0.2cm},
        at={(1.57, 1.03)}, 
        anchor=south, 
        draw=white!80!black, 
        font=\scriptsize
        },
    legend columns=3,
    scale only axis,
    area style,
    xlabel style={font=\footnotesize},
    xlabel={$N_{\rm SS}$},
    xmajorgrids,
    xmin=8,
    xmax=64,
    xtick={8,16,32,64},
    ylabel shift = -1 pt,
    ylabel style={font=\footnotesize},
    ylabel={$T_{\rm SS}$ (ms)},
    ymajorgrids,
    ymajorticks=true,
    ymin=5,
    ymax=160, 
    ymode=log
]

\nextgroupplot[title={\footnotesize a) $\tau=3$ dB.}, 
               every axis title/.style={below, at={(0.5, -0.38)}},
               ytick={5,10,20,40,80,160},
               yticklabels={5,10,20,40,80,160}]

\addplot[fill=mycolor1, draw=black] table[row sep=crcr]{%
8	160\\
16	160\\
32	160\\
64	160\\
}
\closedcycle;
\addlegendentry{$v=1$ m/s}

\addplot[fill=mycolor2, draw=black] table[row sep=crcr]{%
8	80\\
16	80\\
16	160\\
32	160\\
64	160\\
}
\closedcycle;
\addlegendentry{$v=2$ m/s}

\addplot[fill=mycolor3, draw=black] table[row sep=crcr]{%
8	40\\
16	40\\
16	160\\
32	160\\
64	160\\
}
\closedcycle;
\addlegendentry{$v=4$ m/s}

\node[below right, align=left, inner sep=0]
at (rel axis cs:0.561,0.4) {$N^*=3$};
               
\nextgroupplot[title={\footnotesize b) $\tau=7$ dB.}, 
               every axis title/.style={below, at={(0.5, -0.38)}},
               ytick={5,10,20,40,80,160},
               yticklabels=\empty,
               ylabel={}]

\addplot[fill=mycolor1, draw=black] table[row sep=crcr]{%
8	80\\
16	80\\
16	160\\
32	160\\
64	160\\
}
\closedcycle;

\addplot[fill=mycolor2, draw=black] table[row sep=crcr]{%
8	40\\
16	40\\
16  80\\
32  80\\
32	160\\
64	160\\
}
\closedcycle;

\addplot[fill=mycolor3, draw=black] table[row sep=crcr]{%
8	20\\
16	20\\
16  40\\
32  40\\
32  80\\
64	80\\
64	160\\
}
\closedcycle;

\node[below right, align=left, inner sep=0]
at (rel axis cs:0.561,0.4) {$N^*=5.4$};

\nextgroupplot[title={\footnotesize c) $\tau=10$ dB.}, 
               every axis title/.style={below, at={(0.5, -0.38)}},
               ytick={5,10,20,40,80,160},
               yticklabels=\empty,
               ylabel={}]

\addplot[fill=mycolor1, draw=black] table[row sep=crcr]{%
8	20\\
16	20\\
16	40\\
32	40\\
32	160\\
64	160\\
}
\closedcycle;

\addplot[fill=mycolor2, draw=black] table[row sep=crcr]{%
8	10\\
16	10\\
16	20\\
32	20\\
32	80\\
64	80\\
64	160\\
}
\closedcycle;

\addplot[fill=mycolor3, draw=black] table[row sep=crcr]{%
8	5\\
16	5\\
16	10\\
32	10\\
32	40\\
64	40\\
64	80\\
}
\closedcycle;

\node[below right, align=left, inner sep=0]
at (rel axis cs:0.561,0.4) {$N^*=8.8$};

\end{groupplot}
\end{tikzpicture}
	\caption{Feasibility regions for different values of sensitivity threshold $\tau$ and \gls{ue} speed $v$, and for $P_T=18$ dBm.\vspace{-2cm}}
    \label{feasibility_joined}
\end{figure*}

Fig.~\ref{feasibility_joined} depicts the feasibility regions in terms of the upper bounds of parameters $N_{\rm SS}$ and $T_{\rm SS}$ for which the optimization problem in \eqref{OP} is feasible, for $P_T=18$ dBm and $\tau\in\{3, 7, 10\}$~dB. 
These plots were generated using the values in Table~\ref{FR}, and are intended to provide guidelines towards the optimal 5G NR beam management configurations to minimize the energy consumption for RedCap devices.
In general, we observe that as $\tau$ increases, the feasibility regions become smaller. This is because increasing the threshold~$\tau$ translates into a tighter constraint on the SNR (C1 in~\eqref{OP}), thus the optimization problem yields larger values of $N^*$ to increase the beamforming gain. However, this also implies narrower beams, which in turn reduce the angular offset which can be tolerated by the system. 

Furthermore, the size of the feasibility regions is inversely proportional to $v$ and $T_{\rm SS}$, as expected from the analysis in Sec.~\ref{sec3}. Indeed, if the UEs move faster, or if the beam management process takes longer, the angular offset in Eq.~\eqref{theta_k} also increases, and so does the probability that the UEs would lose beam alignment. 
However, we can see from Eq.~\eqref{T_BM} that $T_{\rm SS}$ does not influence the beam management time $T_{\rm BM}$ if $S_D\leq N_{\rm SS}$, i.e., if sending the SSBs requires exactly one burst~\cite{giordani2019standalone}.
Based on the expression of $S_D$ in Eq.~\eqref{eq:SD}, this condition is true if $N_{\rm gNB} \leq   3, 5, 11, 21$, for $N_{\rm SS} = 8, 16, 32, 64$ respectively.
In general, it is convenient to choose $N_{\rm gNB}$ accordingly, to limit the impact of $T_{\rm SS}$ on the shape of the feasibility regions. 


\begin{table}
	\centering
	\renewcommand{\arraystretch}{1.1}
	\caption{Feasibility regions for $P_T=18$~dBm and $\tau\in\{3, 7, 10\}$ dB.}
	\begin{tabular}{ c|l|l|l }
		\hline\
		& \multicolumn{3}{c}{ $vT_{\rm SS}$} \\
		\hline
		\backslashbox{$N_{\rm SS}$}{$\tau$} & $3$ dB &$7$ dB & $10$ dB \\
		\hline
		$8$   &$\leq 0.16$ m  &$\leq 0.08$ m  & $\leq 0.02$ m\\
		$16$   &$\leq 0.72$ m &$\leq 0.16$ m  & $\leq 0.04$ m \\
		$32$   &$\leq 4$ m & $\leq 0.4$ m & $\leq 0.16$ m\\
		$64$   &$\leq 4$ m & $\leq 4$ m & $\leq 0.32$ m\\
		\hline
	\end{tabular}
	\label{FR}
 \vspace{-0.33cm}
\end{table}

\subsection{Energy Consumption}
The feasibility regions in Fig.~\ref{feasibility_joined} show that the smallest (highest) feasible $T_{\rm SS}$ ($N_{\rm SS}$) (i.e., the bottom-right part of the feasibility region) would be the optimal configuration for the beam management. Indeed, this choice implies faster beam alignment and better SNR on average, but also entails the highest overhead as more time resources are used for sending control signals at the expense of data transmissions~\cite[Fig. 17]{giordani2019tutorial}. 
Furthermore, let $\overline{\text{EC}}_{t}$ be the average energy consumption 
for sending SSBs over time, which can be expressed as~\cite{giordani2017improved}:
\begin{align}
	\overline{\text{EC}}_{t}=(P_{\rm UE}T_{\rm SSB}){N_{\rm SS}}/{T_{\rm SS}},
\end{align}
where $P_{\rm UE}T_{\rm SSB}$ represents the average energy consumption for sending one SSB, as per Eq.~\eqref{EC1}.


Overall, 
there exists a trade-off between the beam management periodicity and the resulting overhead and $\overline{\text{EC}}_{t}$, which leads to the optimal values of $T_{\rm SS}$ and $N_{\rm SS}$.
We thus propose to operate in the top-left portion of the feasibility regions, i.e., choosing the highest possible $T_{\rm SS}$ at $N_{\rm SS}=8$. In this way, 
we minimize the average energy consumption per unit time, while still satisfying the SNR constraints as we are in the feasibility regions. For instance, for $P_T=18$ dBm and $\tau\in\{3,7, 10\}$ dB, the optimal configuration for $(N_{\rm SS}, T_{\rm SS})$ at $v=1$ m/s is $(8, 160$ ms$)$, $(8, 80$ ms$)$ and $(8, 20$ ms$)$, respectively. 

\section{Conclusions and Future Work}\label{conclude}
\label{sec:concs}
In this work, we explored the 5G NR beam management design for RedCap devices in an \gls{inf-sh} scenario.
In this scenario, and during beam management, a moving device may lose alignment with the associated beam, potentially resulting in the UE going misdetected. 
Therefore, we formalized an optimization problem to minimize the energy consumption during beam management, while ensuring that some desired \gls{qos} requirements, measured in terms of the misdetection probability, are met. 
Through simulations, we identified the feasibility regions where the problem can be solved, and proposed the optimal values of the beam management parameters for RedCap devices, such as the optimal SSB size and periodicity, to maintain a minimum energy consumption while optimizing latency and overhead.
As part of our future work, we will generalize our optimization problem to other scenarios like smart agriculture, introduce additional mobility models, as well as consider more sophisticated optimization methods, e.g., based on machine learning.
\vspace{-0.5em}
\section*{Acknowledgment}
This work was partially supported by the European Union under the Italian National Recovery and Resilience Plan (NRRP) of NextGenerationEU, partnership on “Telecommunications of the Future” (PE0000001 - program “RESTART”).

\bibliographystyle{IEEEtran}
\bibliography{bibl}

\begin{thebibliography}{10}
\providecommand{\url}[1]{#1}
\csname url@samestyle\endcsname
\providecommand{\newblock}{\relax}
\providecommand{\bibinfo}[2]{#2}
\providecommand{\BIBentrySTDinterwordspacing}{\spaceskip=0pt\relax}
\providecommand{\BIBentryALTinterwordstretchfactor}{4}
\providecommand{\BIBentryALTinterwordspacing}{\spaceskip=\fontdimen2\font plus
\BIBentryALTinterwordstretchfactor\fontdimen3\font minus
  \fontdimen4\font\relax}
\providecommand{\BIBforeignlanguage}[2]{{%
\expandafter\ifx\csname l@#1\endcsname\relax
\typeout{** WARNING: IEEEtran.bst: No hyphenation pattern has been}%
\typeout{** loaded for the language `#1'. Using the pattern for}%
\typeout{** the default language instead.}%
\else
\language=\csname l@#1\endcsname
\fi
#2}}
\providecommand{\BIBdecl}{\relax}
\BIBdecl

\bibitem{atzori2010internet}
L.~Atzori, A.~Iera, and G.~Morabito, ``{The internet of things: A survey},''
  \emph{Computer networks}, vol.~54, no.~15, pp. 2787--2805, May 2010.

\bibitem{zanella2014internet}
A.~Zanella, N.~Bui, A.~Castellani, L.~Vangelista, and M.~Zorzi, ``{Internet of
  things for smart cities},'' \emph{IEEE Internet Things J.}, vol.~1, no.~1,
  pp. 22--32, Feb. 2014.

\bibitem{38875}
3GPP, ``{Study on support of reduced capability NR devices - Rel. 17},'' TR
  38.875, 2020.

\bibitem{varsier20215g}
N.~Varsier, L.-A. Dufr{\`e}ne, M.~Dumay, Q.~Lampin, and J.~Schwoerer, ``{A 5G
  New Radio for Balanced and Mixed IoT Use Cases: Challenges and Key Enablers
  in FR1 Band},'' \emph{IEEE Commun. Mag.}, vol.~59, no.~4, pp. 82--87, May
  2021.

\bibitem{pagin2023nrlight}
M.~Pagin, T.~Zugno, M.~Giordani, L.-A. Dufrene, Q.~Lampin, and M.~Zorzi, ``{5G
  NR-Light at Millimeter Waves: Design Guidelines for Mid-Market IoT Use
  Cases},'' \emph{IEEE ICNC}, 2023.

\bibitem{giordani2019tutorial}
M.~Giordani, M.~Polese, A.~Roy, D.~Castor, and M.~Zorzi, ``{A Tutorial on Beam
  Management for 3GPP NR at mmWave Frequencies},'' \emph{IEEE Commun. Surv.
  Tutorials}, vol.~21, no.~1, pp. 173--196, Firstquarter, 2019.

\bibitem{giordani2017improved}
M.~Giordani and M.~Zorzi, ``{Improved user tracking in 5G millimeter wave
  mobile networks via refinement operations},'' in \emph{Med-Hoc-Net}, 2017.

\bibitem{8613274}
X.~Yang, M.~Matthaiou, J.~Yang, C.-K. Wen, F.~Gao, and S.~Jin,
  ``{Hardware-Constrained Millimeter-Wave Systems for 5G: Challenges,
  Opportunities, and Solutions},'' \emph{IEEE Commun. Mag.}, vol.~57, no.~1,
  pp. 44--50, Jan. 2019.

\bibitem{zhang2018low}
J.~Zhang, L.~Dai, X.~Li, Y.~Liu, and L.~Hanzo, ``{On low-resolution ADCs in
  practical 5G millimeter-wave massive MIMO systems},'' \emph{IEEE Commun.
  Mag.}, vol.~56, no.~7, pp. 205--211, Apr 2018.

\bibitem{sultania2018energy}
A.~K. Sultania, P.~Zand, C.~Blondia, and J.~Famaey, ``{Energy Modeling and
  Evaluation of NB-IoT with PSM and eDRX},'' in \emph{IEEE Globecom Workshops},
  2018.

\bibitem{38869}
3GPP, ``{Study on low-power Wake-up Signal and Receiver for NR -- Release
  18},'' \emph{TR 38.869}, 2023.

\bibitem{3GPPrel16}
------, ``{Study on channel model for frequencies from 0.5 to 100 GHz- Release
  16},'' \emph{TR 38.901}, 2020.

\bibitem{rawat2023optimal}
M.~Rawat, M.~Giordani, B.~Lall, A.~Chaoub, and M.~Zorzi, ``{On the Optimal
  Beamwidth of UAV-Assisted Networks Operating at Millimeter Waves},''
  \emph{IEEE WCNC Workshop}, 2023.

\bibitem{8057288}
W.~B. Abbas, F.~Gomez-Cuba, and M.~Zorzi, ``{Millimeter wave receiver
  efficiency: A comprehensive comparison of beamforming schemes with low
  resolution ADCs},'' \emph{IEEE Trans. Wireless Commun.}, vol.~16, no.~12, pp.
  8131--8146, Dec. 2017.

\bibitem{balanis}
C.~A. Balanis, \emph{Antenna theory: analysis and design}, 4th~ed.\hskip 1em
  plus 0.5em minus 0.4em\relax John Wiley \& Sons, Inc, 2015.

\bibitem{38300}
3GPP, ``{NR} and {NG-RAN} overall description,'' TS 38.300, 2018.

\bibitem{mendez2016}
R.~Méndez-Rial, C.~Rusu, N.~González-Prelcic, A.~Alkhateeb, and R.~W. Heath,
  ``{Hybrid MIMO Architectures for Millimeter Wave Communications: Phase
  Shifters or Switches?}'' \emph{IEEE Access}, vol.~4, pp. 247--267, Jan. 2016.

\bibitem{giordani2018coverage}
M.~Giordani, M.~Rebato, A.~Zanella, and M.~Zorzi, ``{Coverage and connectivity
  analysis of millimeter wave vehicular networks},'' \emph{Ad Hoc Networks},
  vol.~80, pp. 158--171, Nov. 2018.

\bibitem{giordani2019standalone}
M.~Giordani, M.~Polese, A.~Roy, D.~Castor, and M.~Zorzi, ``{Standalone and
  non-standalone beam management for 3GPP NR at mmWaves},'' \emph{IEEE Commun.
  Mag.}, vol.~57, no.~4, pp. 123--129, Apr. 2019.

\end{thebibliography}

\end{document}